\documentclass[prb,preprint,preprintnumbers,amsmath,amssymb,floats]{revtex4}
\usepackage{graphicx}
\usepackage{dcolumn}
\usepackage{bm,epstopdf,soul}

\usepackage{amsmath}
\usepackage{color}
\usepackage{soul}

\begin{document}
\preprint{}

\title{Emergent surface superconductivity of nanosized Dirac puddles in a topological insulator
}

\small
\author{ Lukas Zhao,$^1$ Haiming Deng,$^1$ Inna Korzhovska,$^1$ Jeff Secor,$^1$ Milan Begliarbekov,$^1$ Zhiyi Chen,$^1$
Erick Andrade,$^{2}$ Ethan Rosenthal,$^{2}$ Abhay Pasupathy,$^{2}$ Vadim Oganesyan$^{3,4}$  \& Lia Krusin-Elbaum$^{1,4}$ }
\vspace{3mm}
\affiliation{$^1$Department of Physics, The City College of New York, CUNY, New York, NY 10031, USA }
\affiliation{$^2$Department of Physics, Columbia University, New York, New York 10027, USA}
\affiliation{$^3$Department of Engineering Science and Physics, College of Staten Island, CUNY,  Staten Island, NY 10314, USA}
\affiliation{$^4$The Graduate Center, CUNY,  New York, NY 10016, USA}


\begin{abstract}
\vspace{10mm}

\noindent {\bf
Surfaces of three-dimensional topological insulators have emerged as one of the most remarkable states of condensed quantum matter \cite{Qi2011,Fu2007,Zhang2009,Hsieh2009,Roushan2009} where exotic electronic phases of Dirac particles should arise \cite{Qi2011,Fu2008,Qi2009,Fu2010}. Here we report a discovery of
surface superconductivity in a topological material (Sb$_2$Te$_3$) with resistive transition at a temperature of $\sim 9~\textrm{K}$
induced through a minor tuning of growth chemistry that depletes bulk conduction channels. The depletion shifts Fermi energy towards the Dirac point as witnessed by about two orders of magnitude reduction of carrier density and by very large (${\sim 25,000~ \textrm{cm}^2/\textrm{V}\cdot \textrm{s}}$) carrier mobility.
Direct evidence from scanning tunneling spectroscopy and from magnetic response show that the superconducting condensate forms in surface puddles at unprecedentedly higher temperatures, near 60 K and above. The new superconducting state we observe to emerge in puddles can be tuned by the topological material's parameters such as Fermi velocity and mean free path through band engineering; it could potentially become a hunting ground for Majorana modes \cite{Fu2008} and lead to a disruptive paradigm change \cite{Nayak2008} in how quantum information is processed.
}

\end{abstract}

\maketitle
\normalsize
When nontrivial topological order of the electronic structure is concurrent with spontaneous symmetry breaking associated with strong correlations between particles the outcome is a putative state in which superconducting Cooper pairing does not conform to a conventional view.
Such novel order has been predicted to arise on the surfaces of three dimensional topological insulators, where metallic conduction channels host helical Dirac fermions \cite{Qi2011} that cannot be destroyed by non-magnetic scattering processes and can support unusual electronic phases \cite {Qi2011,Fu2008,Qi2009,Fu2010} when electron correlations are at play.
Up to now, the reported superconducting phases obtained at relatively low temperatures by doping \cite{Hor2010,Sasaki2011,Wray2010} or under very high pressures \cite{Zhang2011,Zhu2013} were found to have increased bulk carrier densities and hence appear to be of bulk origin.

In this report we demonstrate that surface superconductivity in the \textit{p}-type topological material Sb$_2$Te$_3$ can be induced by a small variation in Te vapor pressure during the crystal growth. Tellurium overpressure in a very narrow range, while making no detectable structural changes, acts to compensate bulk carriers so that in the superconducting state the hole carrier density relative to that in the non-superconducting state is reduced, upshifting the Fermi energy from the bulk \textit{valence} bands towards the vicinity of the Dirac point. From frequency dependent magnetic response and local superconducting gap scanning tunneling spectroscopy (STS) we show that superconductivity originates in surface Dirac puddles at remarkably high temperatures, $\gtrsim 60$ K. The superconducting gaps observed in STS can be locally as large as $\sim 25$ meV, which in Bardeen-Cooper-Schrieffer (BCS) theory would correspond to transition temperatures above liquid nitrogen. Global superconducting coherence is reached when interpuddle diffusion of depaired quasiparticles \cite{Spivak2008} establishes a percolative path connecting the puddles, at which point a superconducting resistive transition is observed.

Figure 1a shows resistivity of Sb$_2$Te$_3$ synthesized under $\sim 1.4~\textrm{MPa}$ Te vapor pressure during the high-temperature step of the crystal growth cycle (see Methods) and measured at \textit{ambient} pressure. The system undergoes a transition to zero resistance at the onset temperature $T_{CR} \cong 8.6~K$.  We note that this is the highest $T_C$ observed in a topological material at ambient pressure after synthesis (see Methods, Figs. S1, S2, and Table S1 for chemical and structural analysis).
In the narrow Te pressure range $1.2 < P  < 1.5~\textrm{MPa}$, where superconductivity is found, the hole density is \textit{reduced} by nearly two orders of magnitude (Figs. 1b and S4) to below $ \cong 10^{18}~\textrm{cm}^{-3}$ -- a finding which should be contrasted with the increased electron density recorded in, \textit{e.g.}, superconducting Bi$_2$Se$_3$ doped with Cu \cite{Hor2010}. In the same narrow pressure range the diamagnetism is strongly enhanced (Fig. 1b), but the expected superconducting transition into Meissner field expulsion state is found to take place at a significantly higher temperature $T_{CD} \cong 55~K$, see Fig.~1c.

The diamagnetic transition detected in \textit{dc} magnetization
is sharp; the measured signal is about $10^{-3}$ of the full Meissner value of $1/4\pi$ expected in a superconductor and is nearly flat below $\sim 10~\textrm{K}$. And while resistivity does not display any comparably strong features at $T_{CD}$, scanning tunneling spectroscopy (STS) (Figs.~1d,e and Methods) shows a clear presence of superconducting energy gaps (Fig.~1f and Fig.~S3) that can vary locally from 0 to $\gtrsim 20$ meV. This patchy gap landscape indicates a laterally inhomogeneous superconducting state which within the BCS theory (using the gap equation $2\Delta(0) = 3.5 k_B T_{C}$, where $k_B$ is the Boltzmann constant) has local transition temperatures that can be greater than 60 K, inline with the temperature where the onset of diamagnetic Meissner signal is observed.

The resistive transition downshifts and broadens with increasing external magnetic field (Fig. 2a).  By tracking onset temperature $T_{CR}$ we map the field-temperature $H-T$ phase diagram for the two field orientations: applied along the \textit{c}-axis of the crystal, $H\parallel c$, and parallel to the \textit{ab}-plane, $H\parallel ab$ (upper and lower panels of Fig.~2a, respectively). The low-temperature critical field anisotropy $\sim 1.4$ at first glance is surprisingly small, given the large structural anisotropy of Sb$_2$Te$_3$ (Figs. S1, S2).
This suggests that the limiting field may be of spin rather than orbital origin. We estimate this field \cite{tinkham2012introduction} by comparing magnetic energy ${ \frac {1}{2}}  N(0) g \mu_B^2 H_p^2(0)$ with the superconducting condensation energy at the transition  ${ \frac {1}{2}} N(0) \Delta^2(0)$. Here $N(0)$ is the density of states at the Fermi energy, $g$ is the Land\'{e} $g$-factor, and $\Delta(0)$ and $H_p(0)$ are zero-temperature superconducting gap and upper critical field, respectively.
Using again BCS gap equation and taking the large measured value \cite{Analytis2010} of \textit{g}-factor, $g \cong 50$, gives $H_p(0) = 2.6  {\frac {T_{CR}}{\sqrt{g}}} \approx 3.2~T$, in good agreement with the experiment. We surmise then that depairing is of paramagnetic (Zeeman) origin, with a small anisotropy of the $g$-factor.

Another prominent feature in the $H-T$ phase diagram is the large high-field low-temperature region in the superconducting state \textit{below} the critical field $H_p$ where we observe de Haas-van Aphen (dHvA) quantum oscillations (inset in Fig. 2b and Figs. 4a,b) -- one of the most direct probes of quasiparticle excitations in metals \cite{shoenberg1984magnetic}. This region, bound by the field $H^\star(T)$ at which oscillations first appear is obtained from the temperature dependence of the low-field onset of dHvA.
Occurrence of dHvA oscillations is known in extreme type II superconductors in the vortex state \cite{Maniv2001}, although in conventional superconductors there is a considerable amplitude damping effect below upper critical field $H_{c2}$. The superconducting Sb$_2$Te$_3$ shows no additional damping at the critical field (Fig.~S5). This suggests two possible reasons for robust quantum oscillations in the superconducting state: one is inhomogeneous superconductivity with some residual unpaired fermion quasiparticles 
\cite{Spivak2008}, another is the superconducting gap developing nodes in momentum space \cite{Maniv2001}.

The diamagnetism of the superconducting samples is also highly unusual. Differential diamagnetic magnetic susceptibility $\chi$ (Figs. 3a, b) measured using a relatively low (10~kHz) frequency is strongly field and temperature dependent. Large diamagnetism persists to high ($> 100~\textrm{K}$) temperatures as it smoothly and monotonically reduces to nonsuperconducting values. This should be contrasted with \textit{dc} magnetization (such as shown in Fig.~1b) which has a sharp Meissner transition at $\sim 55~\textrm{K}$.
This unusual differential response is found to vary strongly with frequency of \textit{ac} excitation even for the low ($\sim \textrm{kHz}$) frequencies used (Figs.~3d, e). General analyticity considerations at low external field dictate that low frequency corrections enter quadratically, $\chi(\omega, T) = \chi_0(T) + b(T) \omega^2$, and such behavior is found in resistively shunted Josephson networks and other cases of two fluid (normal + superfluid) dynamics\cite{Hein1991}.
With this in mind we analyzed the frequency dependence of $\chi$ by performing frequency scans at different temperatures, focussing on the low field regime.
The data (inset in Fig. 3g) clearly follows a parabolic frequency dependence.
The fit to this simple form uncovers a spectacular dichotomy between temperature variation of the zero frequency value
$\chi_0(T)$ (Fig.~ 3g and Supplementary Information) and the dispersion coefficient $b(T)$ (Fig. 3f): $\chi_0(T)$ shows a sharp diamagnetic onset in the vicinity of $\sim 55~K$ in close correspondence with the temperature dependence recorded in \textit{dc} magnetization, while the prefactor $b(T)$ smoothly marches toward the near null value at very high temperatures (also
Figs.~S7,~S8 and Supplementary Information Section G).
The zero-frequency response $\chi_0(T)$  is large, much larger than any non-superconducting diamagnetism, and we identify its variation and magnitude with a sharp mesoscopic Meissner transition observed using SQUID (Fig. ~1d), while persistence of the finite frequency response coefficient, $b$, suggests existence of considerable superconducting fluctuations at smaller scales and significantly higher temperatures.

The large separation between $T_{CR}$ and $T_{CD}$ signifies two separate physical processes at work. The patchy network of large local superconducting gaps detected by STS and strong diamagnetic response above $T_{CR}$ unaccompanied by detectable transport signatures is most naturally ascribed to isolated and well separated `puddles' of \emph{local} superconductivity, which we take to be congregating near the surface of the sample and very thin \cite{Park2013}, to account for the observed large orbital anisotropy of the response (Figs. 3g, f), and also the overall depletion of bulk conducting channels (Figs. 1d, 4c). The relatively sharply defined $T_{CD}$ implies that sufficiently many of these puddles are larger than the superconducting coherence length $\xi$ (roughly estimated \cite{WHH1966} at below $\sim 10$ nm) so that we may ignore size effects on the local order parameter. Using Ginzburg-Landau phenomenology \cite{tinkham2012introduction} with $T_{CD}$ being the critical temperature of the puddles, we obtain $\chi\sim 1/\lambda(T)^2\sim T_{CD}-T$ in the strongly type-II limit, valid at high temperatures, in agreement with the approximately linear onset of $\chi_0$ shown in Fig. 3g. Here $\lambda$ is the magnetic penetration depth.

The resistive transition at $T_{CR}$ requires a mechanism for generation of Josephson (phase) coupling and establishment of global coherence of a sufficient fraction of these puddles. We propose that long-lived quasiparticles that give rise to the observed magneto-oscillations also mediate Josephson coupling among puddles \cite{Spivak2008,Eley2011}.
When the separation between puddles is larger than the quasiparticle mean-free path, \emph{and} there is good screening of charge fluctuations in and out of the puddles \cite{Spivak2008}, global coherence is attained once the typical puddle separation, $a$, is comparable to metallic quasiparticles' diffusion length $a\approx L_T=\sqrt{\hbar D/k_B T}\approx \sqrt{\hbar v_F \ell/k_B T}$, where we have used the semiclassical expression for the diffusion constant $D$ in terms of Fermi velocity $v_F$ and mean free path $\ell$, $D\approx v_F \ell$. In this limit $T_{CR}$ is largely controlled by the properties of the metallic matrix and the spatial arrangement of puddles, with the sizes of the puddles, the strength of local order parameter and other details only modifying the criterion above through logarithmic factors \cite{Spivak2008}. The interpuddle spacing can be then estimated from the observed $T_{CR}$ provided the diffusion constant, $D$, of the intervening metal is known.

Next we turn to the analysis of dHvA oscillations, from which we can obtain $D$. Quantum dHvA oscillations (Figs. 2b, 4a,b, and Supplementary Information) are quite remarkable in their own right as they clearly display a beat structure found in two-dimensional quantum well states (2DEG) in semiconductors \cite{Das1989}. Such states have been predicted in topological insulators (TIs) in \textit{ab initio} calculations \cite{Park2013} and detected in angle-resolved photoemission spectroscopy (ARPES) \cite{Bahramy2012} but, to the best of our knowledge, have never been observed in magneto-oscillations. Lifshitz-Kosevich analysis \cite{shoenberg1984magnetic} of these oscillations (see Supplementary Information, Table S2) yields $v_F\simeq5.3\cdot 10^5~\textrm{m/s}$, $\ell\approx 95~\textrm{nm}$, $m=0.065~ m_e$, $k_{F+}=3.7\cdot10^8\ m^{-1}$, $k_{F-}=3.3 \cdot10^8\ m^{-1}$ for Fermi velocity, mean-free path, effective mass, and the two close Fermi wavevectors that induce the beats, respectively.  We note that the overall beat structure closely scales with the magnetic field component transverse to the surface as shown in Fig. 4b for the two field orientations and therefore implies a two-dimensional origin of the signal.
Based on these numbers we obtain the diffusion constant $D=v_F \ell/2 = 0.025~$m$^2$/s, and the typical interpuddle separation $a \approx140~ \textrm{nm}$. Using the actual sample area $A$
to estimate the total number of puddles, and relating the absolute value of the diamagnetic response to the single puddle's response by assuming simply additive contributions of individual monodispersed puddles, we obtain the typical puddle size of about $R\approx 37~ \textrm{nm}$ (see Supplementary Information). It is inline with the size scales of surface Dirac puddles reported in STS studies \cite{Beidenkopf2011}.

Based on this correspondence and on the observed patchy distribution of superconducting gaps, we associate the high temperature superconductivity in Sb$_2$Te$_3$  
with the assembly of Dirac puddles and identify the nature of the metallic matrix that mediates global superconductivity at $T_{CR}$ as 2DEG, the two-dimensional electron gas \cite{Bahramy2012}. One reason for this assignment comes from the spin-orbit splitting $\delta\simeq1.34~\textrm{meV}$ obtained
from the beats in dHvA quantum oscillations (Fig. S6); it is comparable to spin splitting observed in \textit{e.g., }a 2DEG InGaAs/InAlAs heterostructures with strong spin orbit coupling \cite{Das1989}. Figure 4c emphasizes another important reason: in the narrow region of Te overpressure the carrier density is hugely reduced, bringing the Fermi level up to just below the Dirac point. Owing to a peculiar dispersion of the bulk valence bands \cite{Zhang2009} of Sb$_2$Te$_3$, the finite $k$ superconducting pairing spans the $\delta k$ sliver that includes the combined system of Dirac puddles connected via coherent diffusion in the metallic 2DEG as visualized in Fig. 1d. Yet another clue pointing to the surface 2DEG resides in huge carrier mobility
($\sim 25,000 ~\textrm{cm}^2/\textrm{V} \cdot \textrm{s}$)
in the superconducting Sb$_2$Te$_3$, a factor of $\sim 165$ larger than in our nonsuperconducting Sb$_2$Te$_3$ (Table S2) where the Fermi level is pinned deeply within the valence band, or in the related families of TIs where carrier mobilities at best are on the order of a $\sim 1000-3000 ~\textrm{cm}^2/\textrm{V} \cdot \textrm{s}$.

We remark that two-dimensional superconductivity that has been found in 2D electron gas at the interfaces between two \textit{band} insulators, LaAlO$_3$ and SrTiO$_3$ occurs below 200 millikelvin \cite{Reyren2007}, a much lower temperature than we observe; the high $T_C$ in Sb$_2$Te$_3$ is a likely spillover from the superconducting puddles \cite{Nandkishore2013} supporting pairing of helical Dirac holes. Puddles have been known to form at low carrier densities in other two-dimensional Dirac systems \cite{Martin2007} owing to nonlinear screening effects  \cite{dasSarma2007}. The striking new superconducting surface state that emerges in puddles in a TI is potentially tunable through material's control, \textit{e.g.}, of the quasiparticle mean free path, as well as the system's Fermi velocity.

\small

\newpage
\noindent {\textbf{Methods}

\small \noindent Single crystals of Sb$_2$Te$_3$ were grown in evacuated quartz tubes in a horizontal gradient furnace heated to 1000$^o$ C and cooled to room temperature in 7-10 days. The starting materials used were cm-sized chunks of Sb (purity 99.9999\%) and Te (purity 99.9995\%) from Alfa-Aesar used in stoichiometric ratios. The critical parameter was Te pressure during the high temperature segment of the growth process; it was determined from the ideal gas equation $P = nRT_{max}/V$, where $T_{max}$ is the highest temperature used, $V$ is the enclosed volume, $n$ is the umber of moles of the material, and $R$ is the ideal gas constant.
The structure and composition of crystals grown in the 0.4 MPa to 3 MPa pressure range was determined from X-ray diffraction and glow discharge analysis (Evans Analytical Group) that determined impurity content ($< 0.6$ ppm wt) of the final crystals (Supplementary Information). X-ray diffraction was performed in Panalytical diffractometer using Cu K$\alpha~ (\lambda = 1.5405{\AA})$ line from Philips high intensity ceramic sealed tube (3~kW) X-ray source with a Soller slit (0.04 rad) incident and diffracted beam optics. Structural identity was confirmed by micro-Raman spectroscopy of characteristic phonon modes. Carrier densities were determined from Hall resistivity and Shubnikov-de Haas oscillations (Supplementary Information).
Scanning tunneling microscopy (STM) and spectroscopy (STS) measurements were carried out in a homebuilt cryogenic UHV STM at a temperature of 5.8 K. Spectroscopy measurements were performed using a lock-in amplifier running at a frequency of 1.831 kHz and an excitation voltage of 0.2 mV. Spectroscopic imaging was carried out over a grid of points (either $128 \times 128$ or $256 \times 256$ pixels) at various energies using the same lock-in amplifier parameters. Fourier transforms (FFTs) are affine transformed to correct for drift and then hexagonally averaged to enhance signal to noise. Transport and susceptibility measurements were performed in a 14 Tesla Quantum Design PPMS system in 1 mT of He gas. For transport, lithographically patterned Au/Ti contacts were fabricated on 100 nm thin crystals exfoliated onto SiO$_2$/Si substrates. Differential susceptibility was measured in a compensated pickup-coil detection configuration with the excitation/detection coils designed to align with the applied static field. The \textit{ac} excitation amplitude was set at $10^{-5}~\textrm{T}$ in a frequency range up to 10 kHz. The system was calibrated using paramagnetic Pd standard and superconducting Nb. \textit{dc} magnetization measurements were performed using Quantum Design SQUID Magnetometer in up to 5.5 Tesla fields.

 \small


\small

\vspace{2mm}

\normalsize
\vspace{5mm}
\noindent \textbf{Acknowledgements:} We are grateful to David Huse, Steven Kivelson, Boris Spivak, Kyungwha Park, Kamran Behnia, Steven Girvin, and Shivaji Sondhi for enlightening discussions. This work was supported in part by NSF under DMR-1122594 (LK) and DMR-0955714 (VO), and by DOD-W911NF-13-1-0159 (LK).

\noindent\section*{FIGURE LEGENDS}

\noindent \textbf{Figure 1 $\mid$ Superconductivity in a topological insulator Sb$_2$Te$_3$.}
\textbf{a,} Resistivity of an exfoliated 100 nm thin  Sb$_2$Te$_3$ crystal synthesized under $\sim 1.4~\textrm{MPa}$ Te vapor pressure (see Methods)
shows onset of transition to zero resistance at $T_{CR} \cong 8.6~K$. Inset shows Hall contact configuration used. \textbf{b,} Diamagnetic susceptibility (left) at 1.9 K measured in a 0.2 T field shows huge diamagnetism (Meissner effect) only in the narrow Te vapor pressure range. It is anticorrelated with the measured carrier density (right) which in the superconducting state is \textit{decreased} by a factor of nearly 100. Outside this range samples are nonsuperconducting. \textbf{c,} \textit{dc} magnetization measured at 0.2 Tesla in SQUID shows a sharp diamagnetic onset at $T_{CD} \approx 55~K > T_{CR}$. The data taken under zero-field-cooled (ZFC) and field-cooled (FC) conditions are essentially identical, consistent with negligible vortex pinning in 2D (ref. [\onlinecite{tinkham2012introduction}]). \textbf{d,} Illustration of superconductive puddles (blue, size $2R$) of Dirac bands, where pairing occurs at high temperature ($T_{CD}$), connected to the 2DEG metallic matrix (grey) which establishes a percolative path (dashed red line) at $T_{CR}$. The electronic Dirac dispersion and spin-orbit split dispersion of 2DEG are also sketched, see text. \textbf{d,e,f,} STM and STS scans were performed in areas $\sim 25$ nm each, spaced $\sim 200$ nm apart. \textbf{e,} A typical topograph of a scanned area shows well ordered hexagonal lattice, with differential conductance $\textrm{d}I/\textrm{d}V$ (from the average of 500 scans) shown in \textbf{f}. Depending on the scan area, the gaps $2\Delta$ vary from zero to  $\gtrsim 20$ meV, see Fig.~S3.

\vspace{5mm}

\noindent \textbf{Figure 2 $\mid$ Magnetic field dependence of the superconducting transition in Sb$_2$Te$_3$.}  \textbf{a,} Resistive transition temperature downshifts with increasing magnetic field applied (top) along the \textit{c}-axis of the crystal, $H\parallel c$, and  (bottom) parallel to the \textit{ab}-plane, $H\parallel ab$.  The onset of superconductivity is indicated by the arrows. \textbf{b,} $H-T$ phase diagram of the superconducting state has relatively small zero-temperature anisotropy of $\sim 1.4$. The critical field values at $T \rightarrow 0$ agree with the paramagnetic (Zeeman) depairing field  $H_p$, see text. Inset: de Haas van Alfen quantum oscillations (dHvA) shown for $H\parallel c$  at 1.9 K persist below $H_p$ down to the temperature dependent onset field $H^\star$.

\vspace{5mm}

\noindent \textbf{Figure 3 $\mid$ Unusual high temperature diamagnetism in the superconducting Sb$_2$Te$_3$.}
\textbf{a,} Differential magnetic susceptibility $\chi$ \textit{vs}. temperature for several values of magnetic field $\mu_0H \parallel c$-axis. Meissner-like signal is `flat' below $T_{CR}$ and monotonically vanishes at much higher ($> 100~\textrm{K}$) temperatures. $\chi$ was measured by applying a small \textit{ac} excitation field  $h_{ac} = 10^{-5}~\textrm{T}$ at $f = 10~\textrm{kHz}$. \textbf{b,} Full temperature and field dependence of $\chi$. \textbf{c,} $\chi$ \textit{vs}. magnetic field $\mu_0H \parallel c$-axis for a series of temperatures. The pronounced dHvA oscillations are apparent at 1.9 K. \textbf{d,} $\chi$ strongly depends on frequency. \textbf{e,} Full frequency and temperature dependence of $\chi$.
Inset in \textbf{g}: Analysis of the frequency dependence of $\chi$ shows it to be quadratic in $\omega = 2\pi f$ (see text).
\textbf{f,} The prefactor $b(T)$ in the $\omega^2$ term dominates the total variation of $\chi$ at finite frequencies. This variation is consistent with kinetic inductance of the patchy distributed 2DEG network discussed in the text. \textbf{g,} Main panel: The zero frequency $\chi_0(T)$ obtained from the fits to $\chi(T) = \chi_0(T) + b(T) \omega^2$ shows a sharp onset at the same $T_{CD}$.

\vspace{5mm}

\noindent \textbf{Figure 4 $\mid$ Signatures of 2D superconductivity in Sb$_2$Te$_3$.}
\textbf{a,} dHvA oscillations show beats arising from two very close oscillation periods. Inset: Fast Fourier transform (FFT) of the signal. The beats are a signature of spin-splitting by a strong spin-orbit interaction in 2DEG surface regions \cite{Bahramy2012}. \textbf{b,} dHvA oscillations \textit{vs.} inverse transverse component of magnetic field $H_\perp = H\textrm{cos} \theta$ scale with $H_\perp$. \textbf{c,} Hole carrier density $n$ \textit{vs.} Te pressure in the superconducting Sb$_2$Te$_3$ (in the narrow $\Delta P$ vicinity of $\sim 1.4$ MPa of Te pressure) and non-superconducting Sb$_2$Te$_3$ (outside $\Delta P$ ) states. In the superconducting region $n$ is reduced, bringing the chemical potential from the deep inside the bulk valence band to just below the Dirac point, as illustrated in sketches of the band structures in all three regions. The conductivity of the superconducting Sb$_2$Te$_3$ remains hole-like (\textit{p}-type), with Fermi level crossing both, the Dirac bands and a 2DEG sliver of the valence band (upper middle sketch).

\newpage

\hspace{-25mm}
\includegraphics[width=20cm]{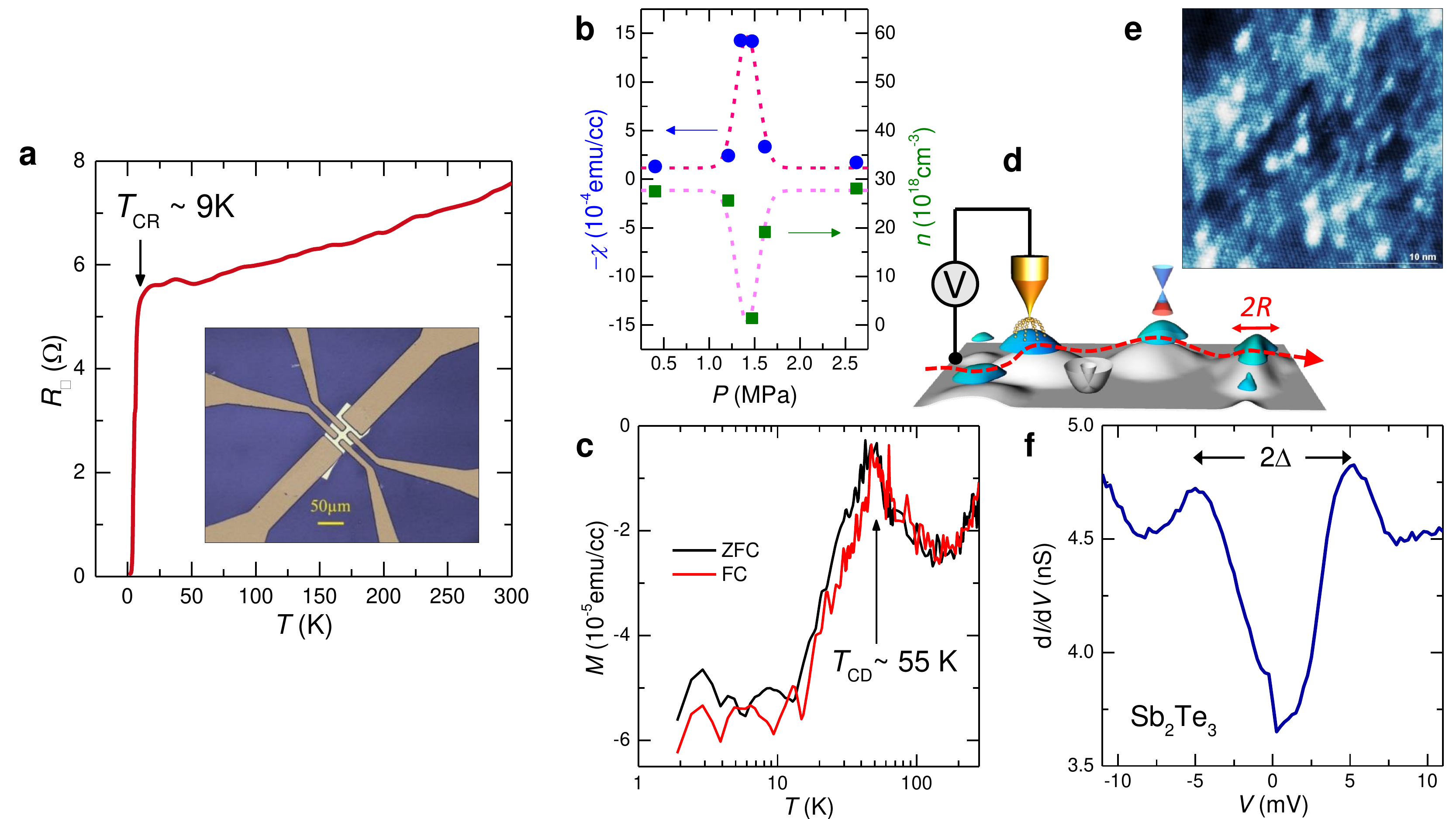}%
\vfill\hfill Fig.~1 LZ  \eject

\hspace{-30mm}
\includegraphics[width=22cm]{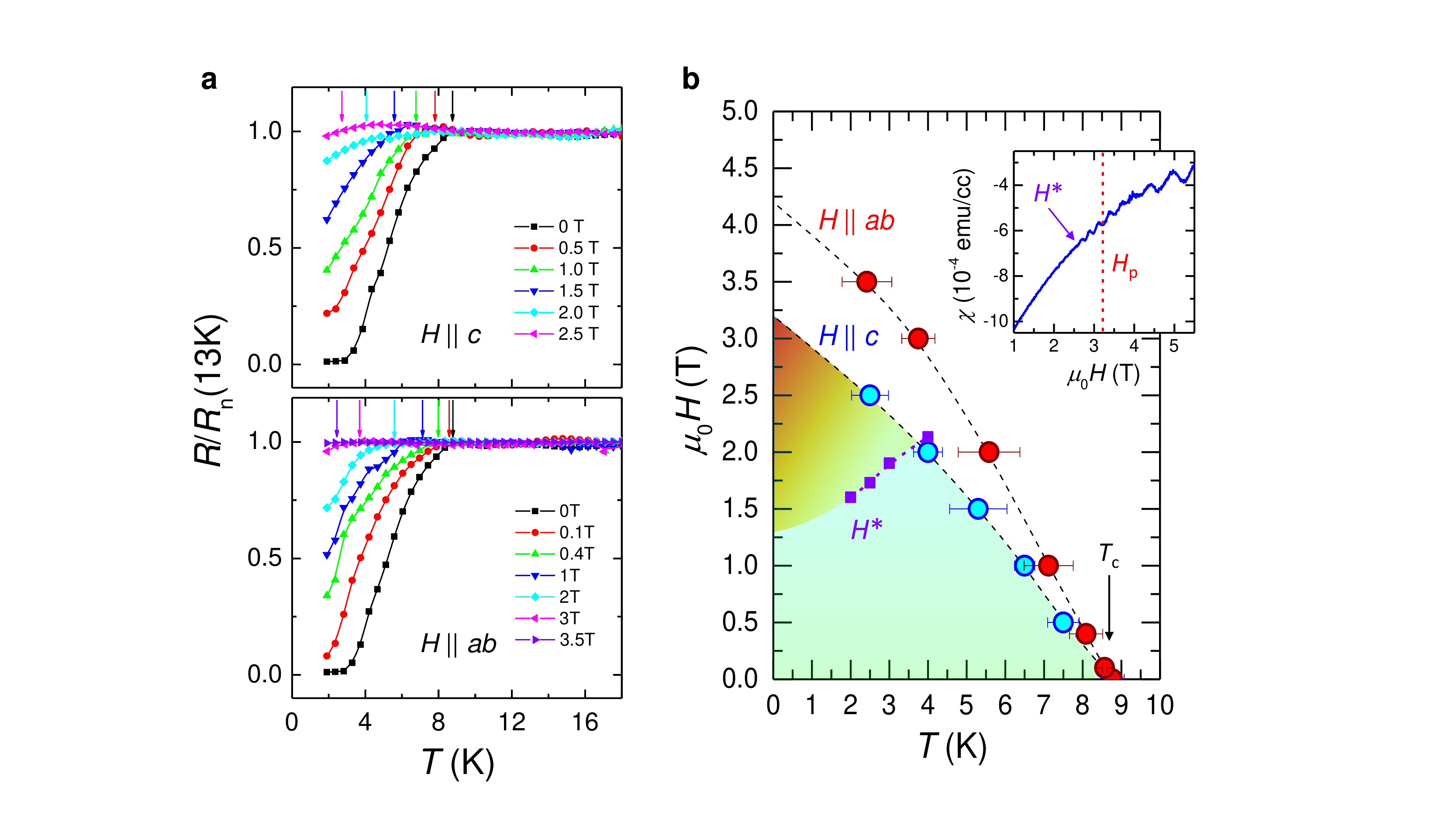}%
\vfill\hfill Fig.~2 LZ  \eject

\hspace{1mm}
\includegraphics[width=14cm]{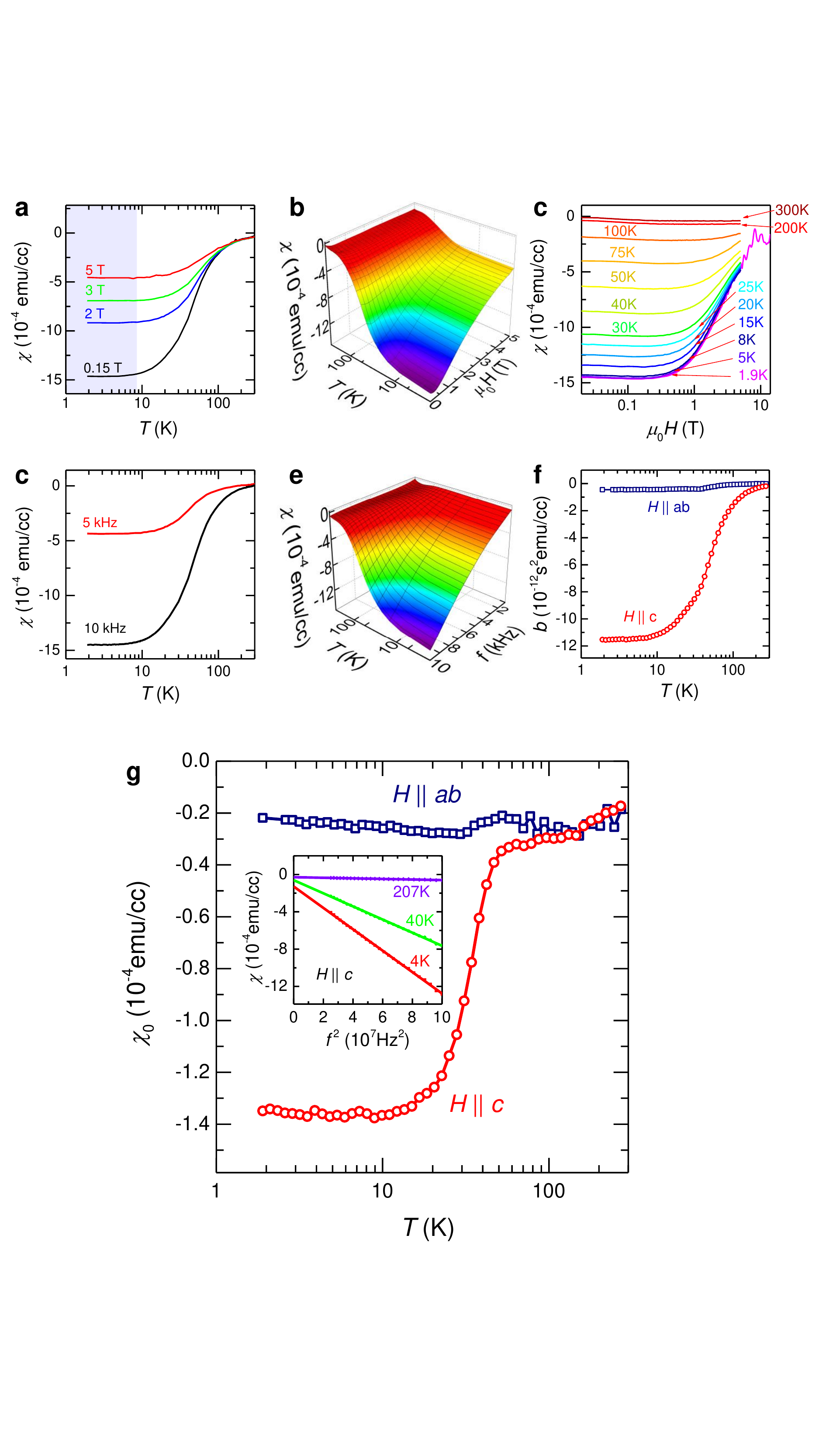}%
Fig.~3 LZ  \eject

\hspace{-35mm}
\includegraphics[width=24cm]{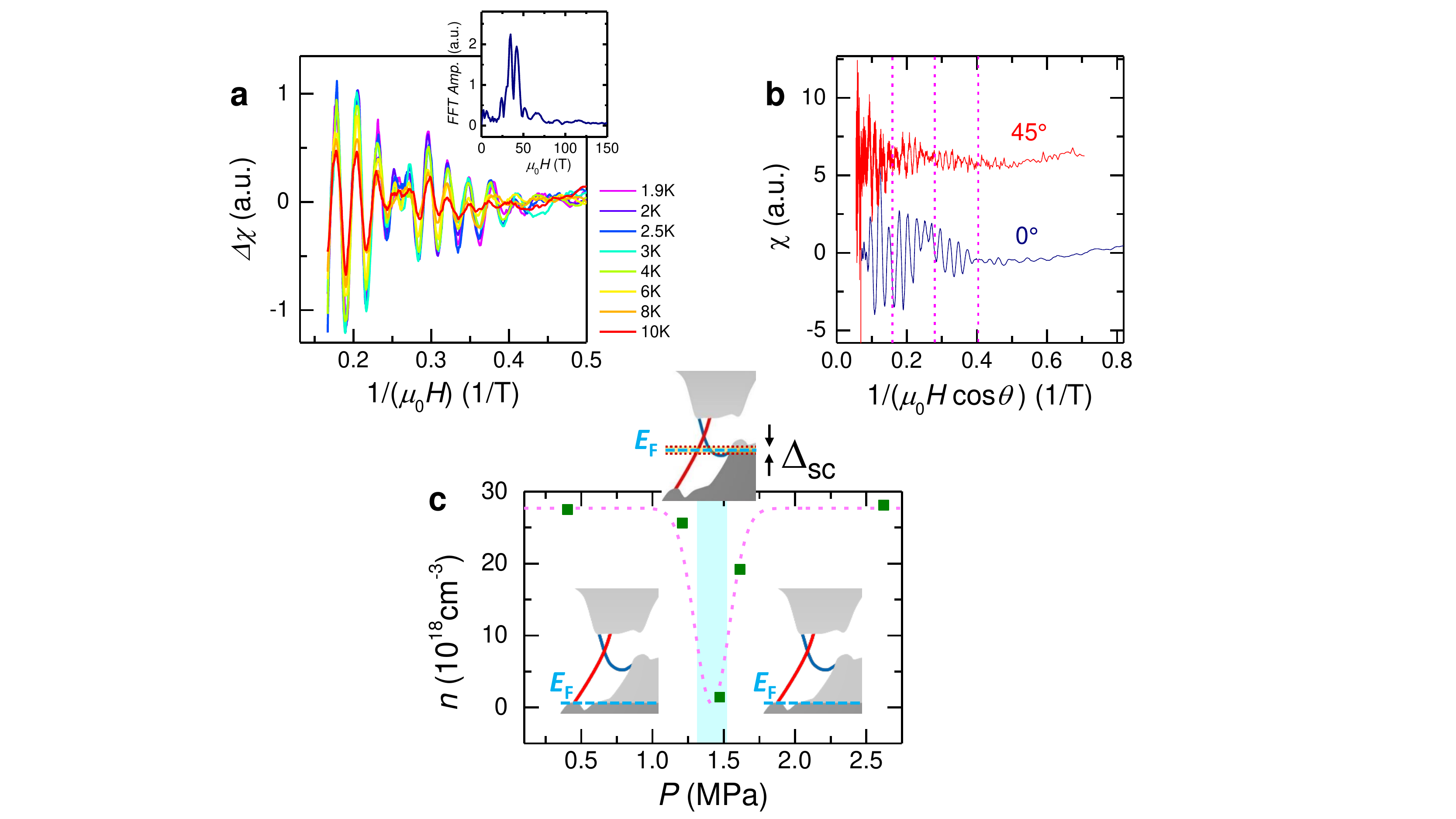}%
\vfill\hfill Fig.~4 LZ  \eject

\newpage

\begin{center}
\large \textbf{Supplementary Information: \\
Emergent surface superconductivity of nanosized Dirac puddles in a topological insulator} \\
\end{center}

The Supplementary Information is organized into seven sections:\\

\noindent
\\
\textbf{(A)} Structural analysis (X-ray, Raman, TEM)\\
\textbf{(B)} Elemental characterization\\
\textbf{(C)} Superconducting gap mapping using scanning tunneling spectroscopy\\
\textbf{(D)} Variation of carrier density measured via Hall and de Haas-van Alphen~ effects\\
\textbf{(E)} Lifshitz-Kosevich analysis and determination of spin-orbit splitting\\
\textbf{(F)} Estimating interpuddle separation\\
\textbf{(G)} Frequency and temperature dependence in the inductive linear response

\maketitle

\clearpage

\makeatletter 
\renewcommand{\thefigure}{S\@arabic\c@figure}
\makeatother
\section*{A. Structural analysis}
\begin{figure}[h!]
\vspace{-20mm}
\begin{center}
\vspace{10mm}
\includegraphics[width=\linewidth]{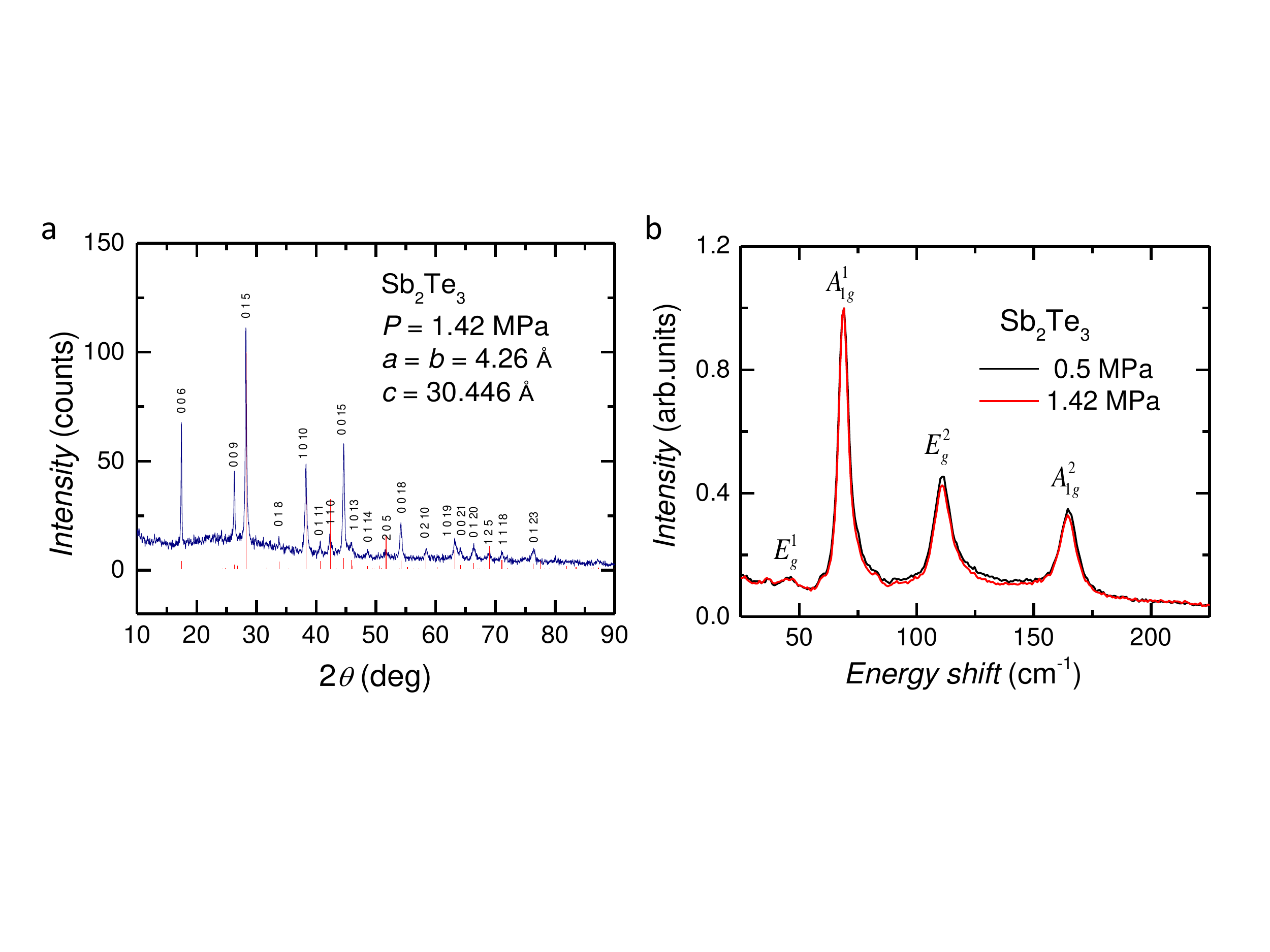}
\vspace{-30mm}
\end{center}
\caption{\small (a) X-ray diffraction spectrum for Sb$_2$Te$_3$ grown in a sealed quartz tube under Te vapor pressure $P = 1.42~ \textrm{MPa}$ collected in Panalytical diffractometer using Cu K$\alpha~ (\lambda = 1.5405{{\AA}})$ line from Philips high intensity ceramic sealed tube (3~kW) X-ray source with a Soller slit (0.04 rad) incident and diffracted beam optics. Rietveld refinement lines shown in red are in full correspondence with the measured spectra. The obtained lattice parameters $a = b = 4.26~{{\AA}}$ and $c = 30.46~{{\AA}}$  remain unchanged in the 0.5 - 2~MPa pressure range, indicating that in this pressure range there is no structural change. (b) The robustness of the structure is also apparent in the identical micro-Raman spectra for superconducting (red) a non-superconducting (black) Sb$_2$Te$_3$.  The \textit{E} phonon modes are doubly degenerate modes in the \textit{ab}-plane and the \textit{A} modes are nondegenerate vibrations with atomic motion along the \textit{c}-axis.  The spectra were taken in ambient conditions in a backscattering geometry with linearly polarized excitation in the ab plan and normalized to the out of plane vibration at $70~\textrm{cm}^{-1}$.}
\protect
\label{fig:XRD}
\end{figure}
\normalsize
Transmission Electron Microscopy (TEM), X-ray diffraction and micro-Raman spectra of superconducting Sb$_2$Te$_3$ indicate absence of any structural changes induced by Te overpressure.
We confirmed that the relatively low Te vapor pressure during the synthesis did not alter the layered rhombohedral van der Waals structure or lattice parameters of Sb$_2$Te$_3$ as determined from the X-ray diffraction (XRD) spectra (Figs. S1 and S2).
\begin{figure}[h!]
\vspace{-5mm}
\includegraphics[width=\linewidth]{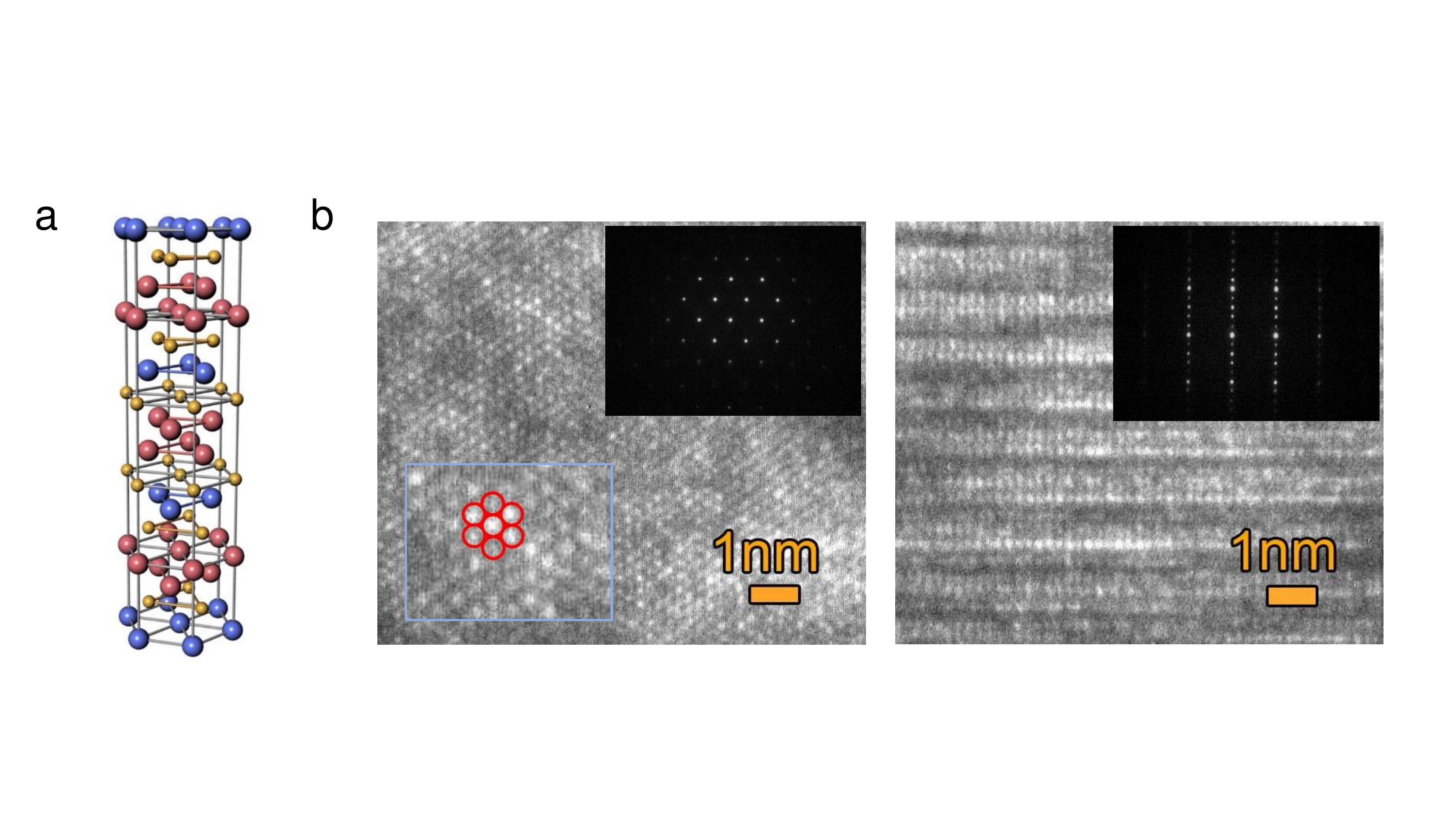}
\vspace{-20mm}
\caption{\small (a) Rhombohedral layered structure of the crystal constructed from the X-ray diffraction (XRD) spectra (see Fig.~S1). The structure remains van der Walls type, with three quantuple layers per unit cell and with the XRD-determined lattice parameters up to $\sim 2-3~\textrm{MPa}$. (b) High resolution transmission electron microscopy images of the crystal are consistent with the XRD. A hexagonal lattice in the $ab$-plane is shown on the left and a layered van der Walls structure along the $c$-axis (normal to the
(00\={1}) cleavage plane is shown on the right.}
\protect
\label{fig:TEM}
\end{figure}

\section*{B. GDMS Elemental analysis}

\baselineskip24pt
Glow discharge mass spectrometry (GDMS) analysis of superconducting  Sb$_2$Te$_3$ lists the impurity content in these crystals, see Table S1. The impurity content is the same in the non-superconducting Sb$_2$Te$_3$. GDMS was performed by Evans Analytical Group (EAG). It has detection limits on the sub-ppm range for most elements that are nearly matrix-independent. In this technique, collisions between the gas-phase sample atoms and the plasma gas pass energy to the sample atoms, exciting the atoms. The atoms then loose their energy by emitting light with the atom specific wavelength. From the intensity of emitted light the atomic concentration can be determined. Sample atoms are also ionized through collisions that then are detected by mass spectrometry. The impurity content of all elements detected in Sb$_2$Te$_3$ matrix is less that a small fraction of ppm.

\begin{figure}[h!]
\begin{center}
\includegraphics[width=15cm]{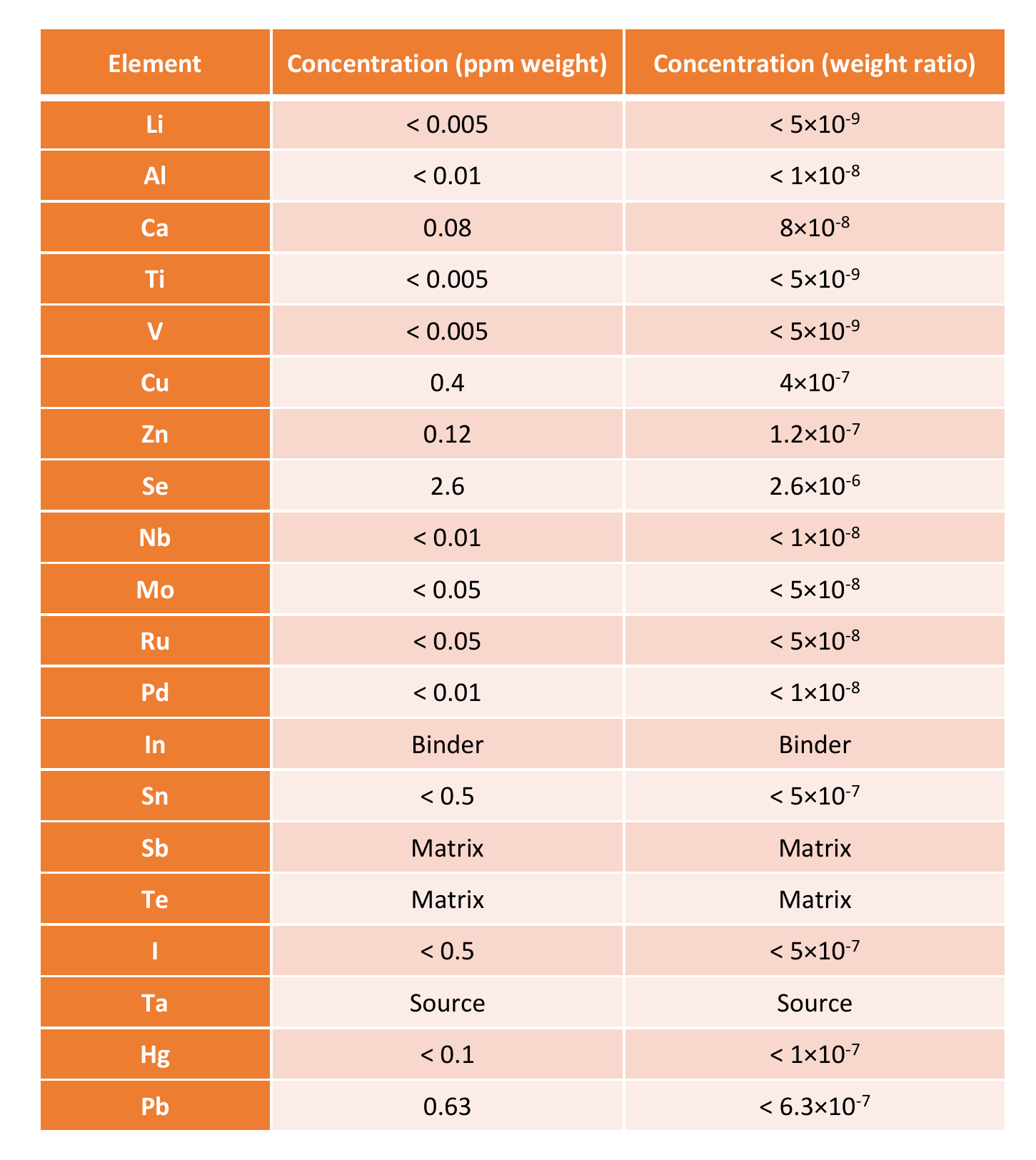}

\textbf{TABLE S1}
\end{center}
\protect
\end{figure}

\normalsize

\clearpage
\section*{C. Superconducting gap mapping using scanning tunneling spectroscopy}
\begin{figure}[h!]
\vspace{-5mm}
\hspace{-100mm}
\begin{center}
\includegraphics[width=16cm]{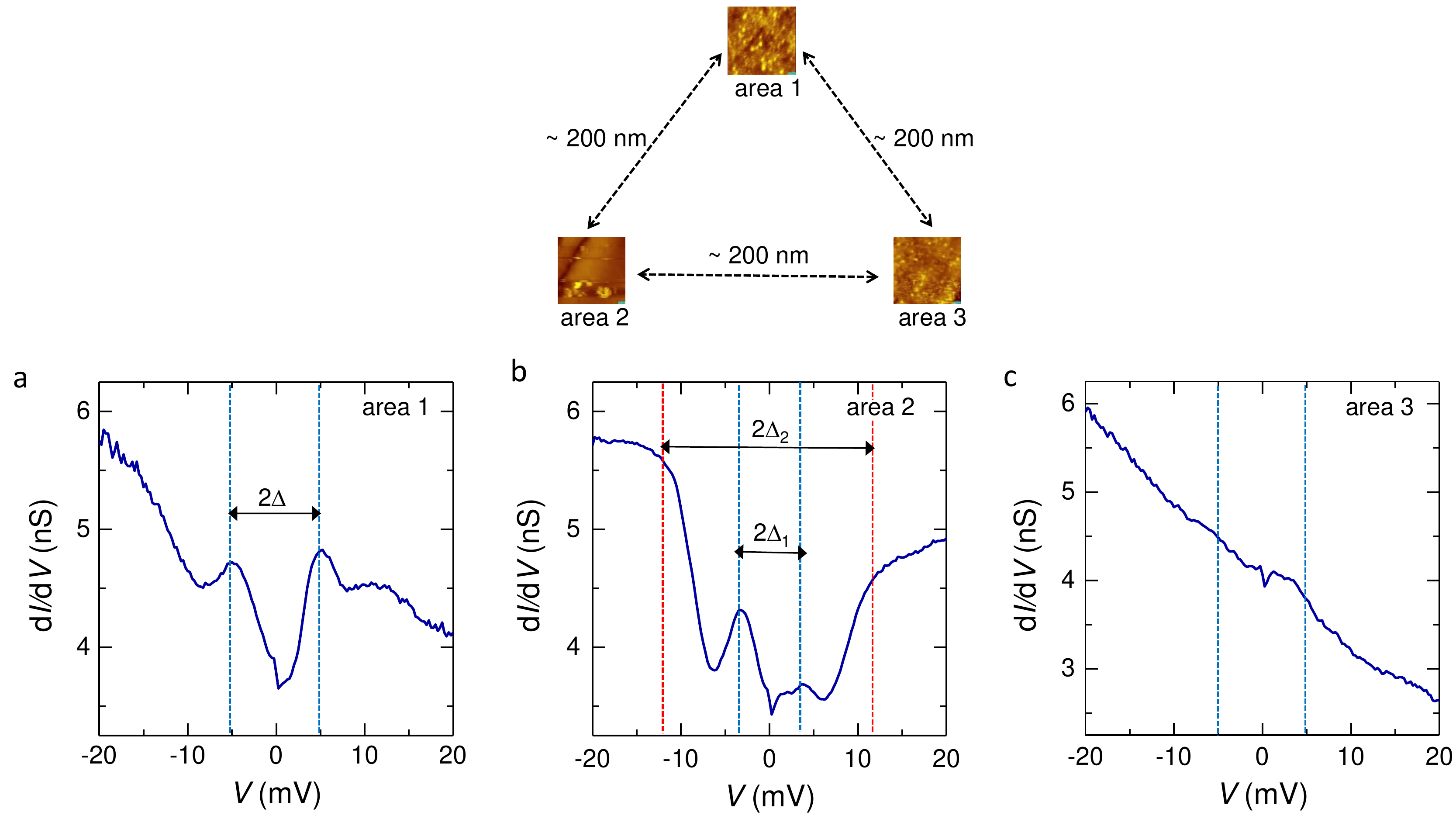}
\vspace{-5mm}
\end{center}
\caption{\label{fig:STM}
\small Scanning tunneling spectroscopy from superconducting surfaces of Sb$_2$Te$_3$. The scans in (a)-(c) were performed in three different surface areas about 25 nm each, spaced $\sim$ 200 nm apart, as sketched in the outset. (a) Differential conductance $\textrm{d}I/\textrm{d}V \textrm{(}V\textrm{)}$ was obtained from the average of 500 scans. In area 1 it shows well articulated coherence peaks and the gap $2\Delta \approx 10~\textrm{meV}$ corresponding to $T_C \sim 30~\textrm{K}$ estimated from BCS gap equation. (b) $\textrm{d}I/\textrm{d}V$ in area 2 has a more complex behavior with two gaps evident: a smaller one $2\Delta_1 \approx 8~\textrm{meV}$ and a larger one $2\Delta_2 \gtrsim 20~\textrm{meV}$ corresponding to $T_C \gtrsim 70~\textrm{K}$. (c) $\textrm{d}I/\textrm{d}V$ in area 3 shows essentially no gap. The slope in measured $\textrm{d}I/\textrm{d}V$ in area 3 is also seen in good metals such as copper \cite{Eigler1993} or gold \cite{Crommie1998} where it is due to band structure effects. Other samples grown in the Te pressure range where superconductivity is found show similar distribution of gap energies, consistent with superconducting puddles with average $T_{CD} \sim 55~\textrm{K}$ embedded in the gapless matrix gleaned in area 3.
}
\protect
\end{figure}
\normalsize
\baselineskip24pt
The variation in the articulation of coherence peaks as well as finite density of states (DOS) at low energies is commonly seen in strongly correlated systems such as high-$T_C$ cuprates \cite{Davis2005} and heavy fermion superconductors \cite{HeavyFermion1999} where it is a consequence of electronic inhomogeneity and strong Coulomb repulsion.
It can also be detected when the Fermi energy is small as in the case of superconductivity at the LAO/STO interfaces \cite{Mannhart2013}.
\normalsize

\clearpage
\section*{D. Hall resistivity and de Haas van Alfen (dHvA) quantum oscillations in superconducting Sb$_2$Te$_3$}
\begin{figure}[h!]
\vspace{-10mm}
\hspace{-5mm}
\begin{center}
\includegraphics[width=15cm]{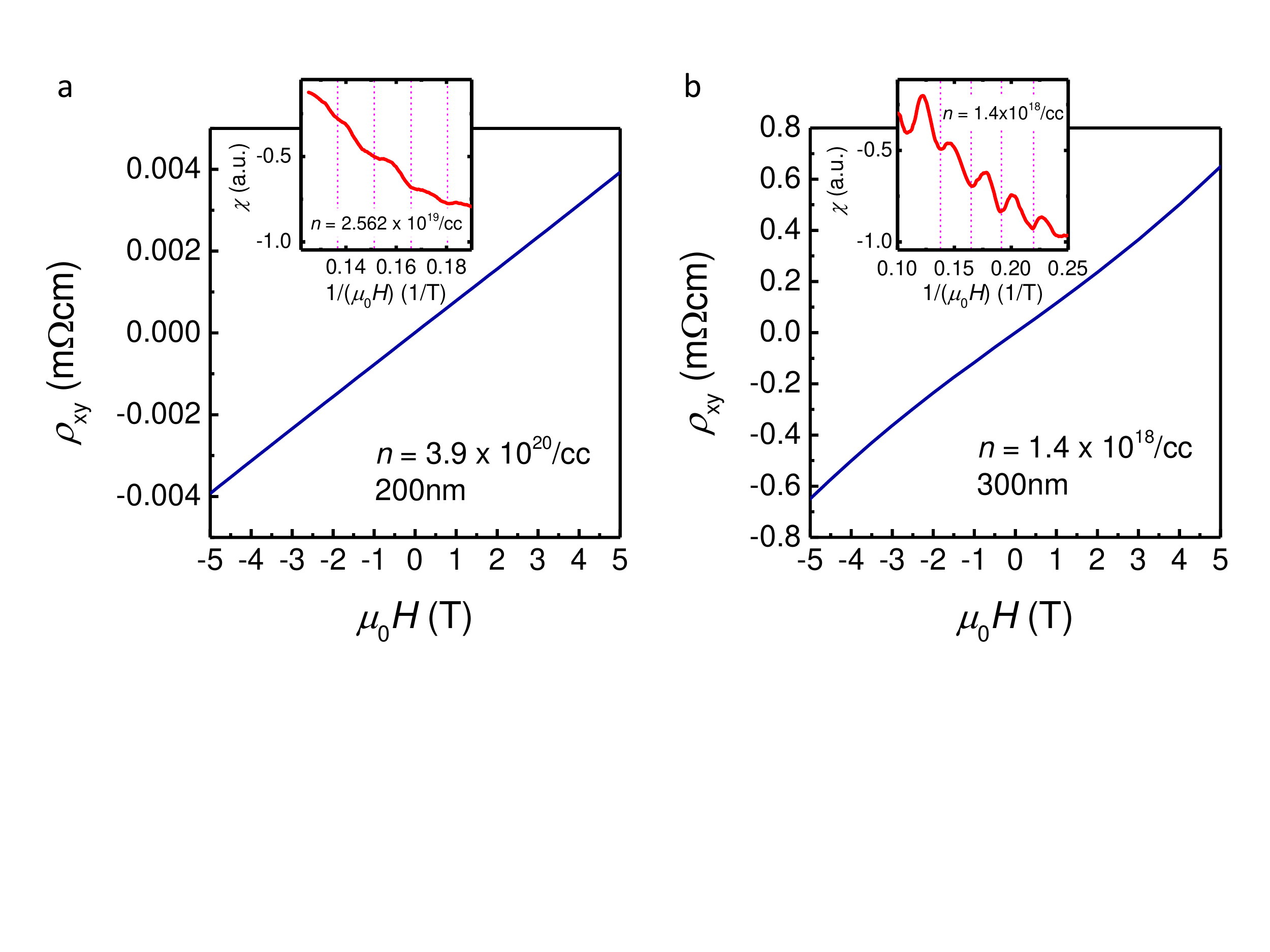}
\vspace{-30mm}
\end{center}
\caption{\label{fig:Hall}
\small (a) Main panel: Hall resistivity of non-superconducting Sb$_2$Te$_3$ shows it to be \textit{p}-type. Carrier density determined from Hall is $n = 3.9 \times 10^{20}/\textrm{cc}$. However, carrier concentration determined from dHvA is $n = 2.56 \times 10^{19}/\textrm{cc}$, over an order of magnitude lower. Similar differences are found in other TIs when chemical potential is located deeply inside either valence or conduction bands. This is reconciled when taking into account the incoherent addition of contributions to Hall conductivity from hexagonal ``pockets" (factor of 6), while the Fermi cross-sections in dHvA are sampled coherently. The additional factor of 2 comes from spin-splitting in the bulk bands, fully accounting for the differences in $n$. (b) Main panel: Hall resistivity of superconducting Sb$_2$Te$_3$ shows it also to be \textit{p}-type. Here, however, carrier densities determined from Hall and dHvA (top outset) are identical $n = 1.4 \times 10^{18}/\textrm{cc}$, and over an order of magnitude lower than in the superconducting samples. This is consistent with the location of the Fermi level just below the Dirac point, and nearly on top of the valence band (Fig. ~4c).
}
\protect
\end{figure}
\normalsize
\baselineskip24pt
Variation of carrier density with Te pressure (Fig. 1 in the main text) is obtained by Hall effect and dHvA oscillations as shown here for two samples. We note that all Sb$_2$Te$_3$ crystals are \textit{p}-type (the charge carriers are holes). The non-superconducting crystals synthesized at Te vapor pressures $P < 1.2~\textrm{MPa}$ and $P> 1.55~ \textrm{MPa}$ have metallic-like temperature dependence ($\textrm{d}R/\textrm{d}T > 0$) of resistance and carrier densities $n \sim 2 \times 10^{19}-10^{20}~\textrm{cm}^{-3}$. In the superconducting Sb$_2$Te$_3$ crystals ($P\sim 1.4~\textrm{MPa}$) the carrier density is over an order of magnitude lower.
\normalsize
\section*{E. de Haas van Alfen oscillations: Lifshitz-Kosevich analysis; spin-orbit splitting}

\begin{figure}[h!]
\hspace{-20mm}
\includegraphics[width=18cm]{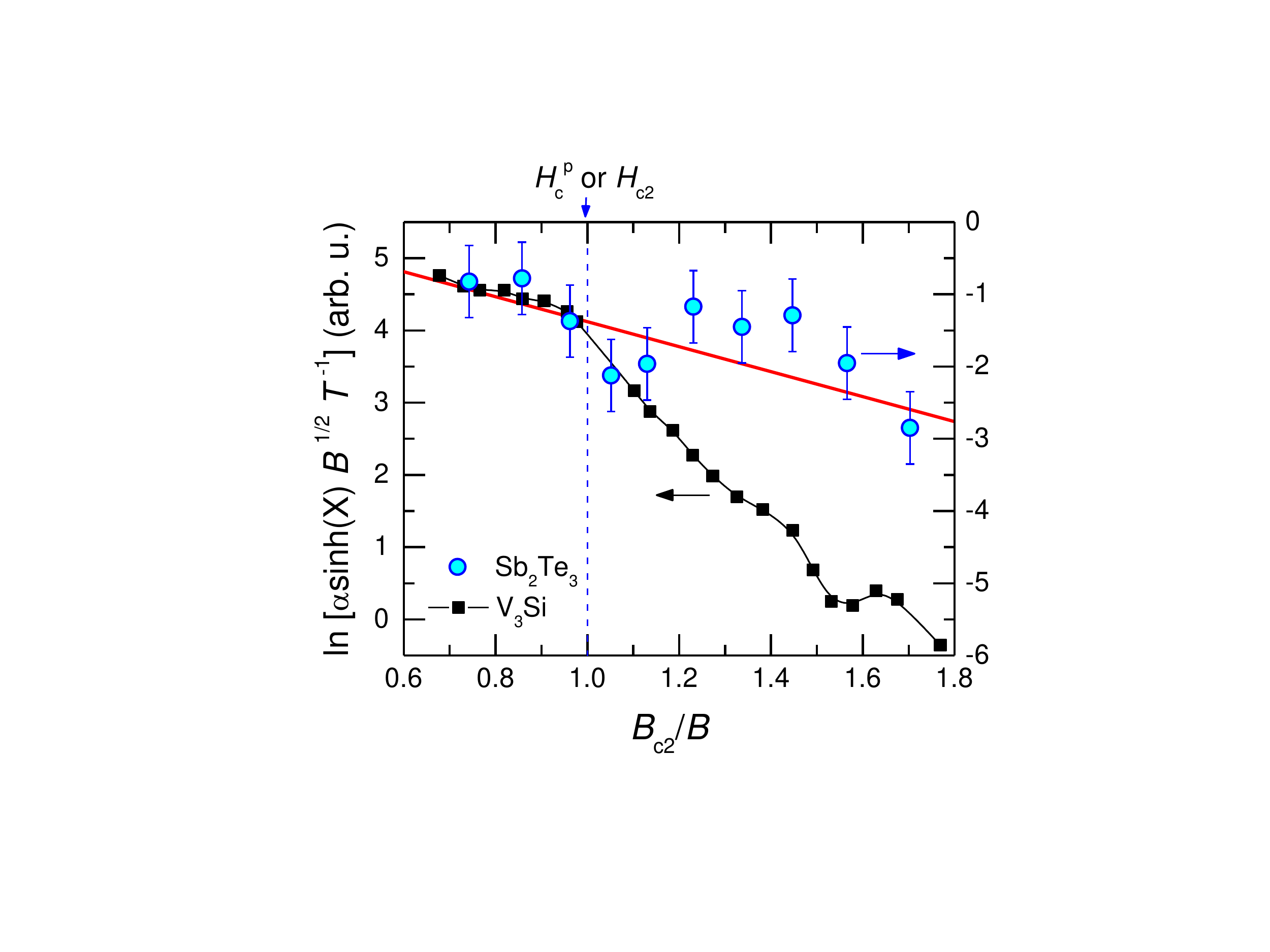}
\vspace{-20mm}
\caption{\label{fig:Dingle}
\small Comparison of de Haas van Alfen (dHvA) oscillation amplitude damping for V$_3$Si and Sb$_2$Te$_3$. Plot shows field dependence of
$D =\textrm{ln} [\alpha \textrm{sinh}(X) B^{1/2} T^{-1}]$ (known as Dingle plot) that shows the change in the dHvA oscillation amplitude upon crossing the superconducting limiting field. Here $\alpha (T,B)$ is the amplitude of quantum oscillations, $X=2\pi^2 k_{B}T/\hbar \omega_c$ and $\omega_c = eB/m_c$ is the cyclotron frequency. The data shown for V$_3$Si are from Ref.~\cite{Corcoran1994} where the field scale was normalized to upper critical field $B_{c2}$. In conventional superconductor V$_3$Si there is an additional attenuation of the oscillation amplitude at the transition into the superconducting state observed in many extreme type II superconductors \cite{Maniv2001}. In a nontrivial superconducting Sb$_2$Te$_3$ there is no additional damping at the superconducting limiting field $H_p$. Here we shifted the Dingle scale (in the same units) to overlay the data for both systems in their normal states (red line). The observed modulation of the Dingle factor $D$ in Sb$_2$Te$_3$ is a result of beats in dHvA. The origins of the absence of attenuation can be of two sources: (i) inhomogeneous (puddle) superconductivity, and (ii) the superconducting order parameter is nodal \cite{Maniv2001}.
}
\protect
\end{figure}

\begin{figure}[h!]
\vspace{-45mm}
\includegraphics[width=16cm]{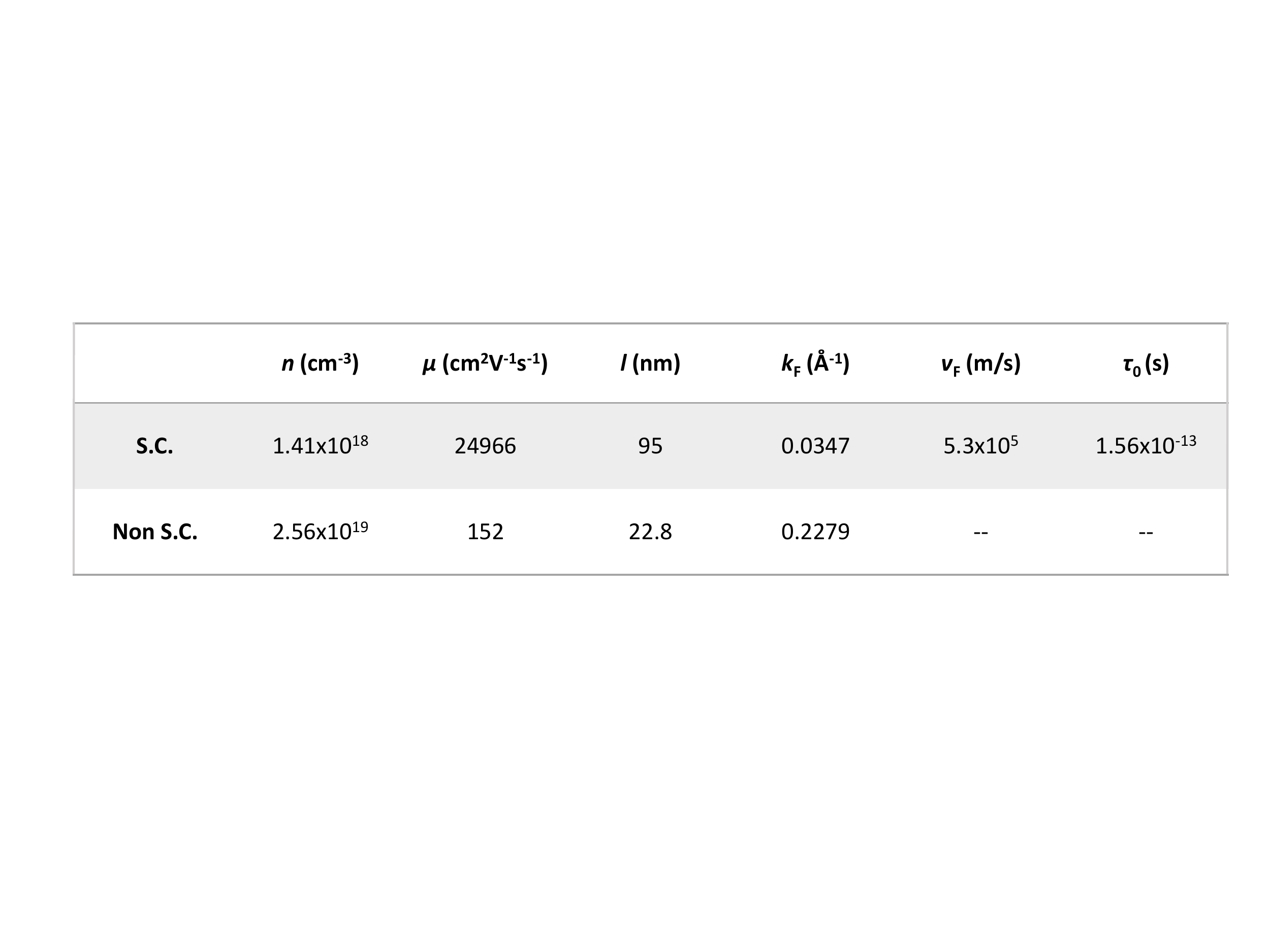}
\begin{center}
\vspace{-40mm}
\textbf{TABLE S2}
\end{center}
\end{figure}
\protect

\begin{figure}[h!]
\begin{center}
\includegraphics[width=16cm]{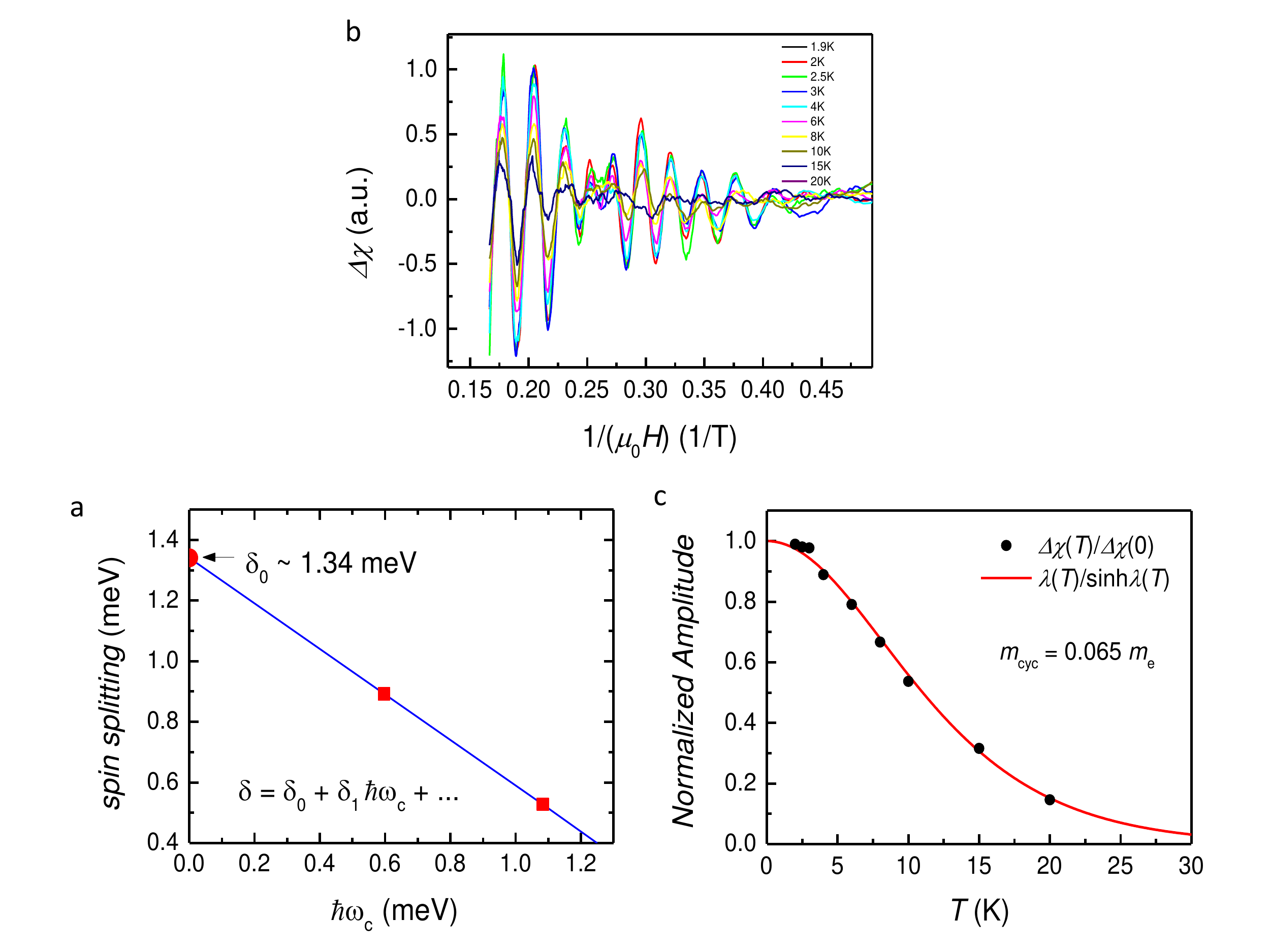}
\caption{\label{fig:SOC}
\small (a) Determination of spin-splitting due to spin-orbit coupling from (b) the beats in de Haas van Alfen quantum oscillations in superconducting Sb$_2$Te$_3$. (c) Determination of cyclotron mass from the temperature dependence of dHvA oscillations in (b).
}
\end{center}
\end{figure}
\protect


\baselineskip24pt
Electronic parameters of Sb$_2$Te$_3$, such as cyclotron mass, carrier densities, or Fermi velocities are obtained from de Haas van Alfen (dHvA ) oscillations using Lifshitz-Kosevich theory \cite{LIFSHITS1956}, see Figs S2-S4, and Table S2. The most striking aspect of the measured dHvA signal is the absence of any significant amplitude damping upon entering the superconducting state   shown in the Dingle plot \cite{Dingle1952,Janssen1998} in Fig. S3.

The beating effect in dHvA oscillations implies the existence of two closely spaced frequency components with similar amplitudes (see Fig.~4a). This has been seen in In$_x$Ga$_{1-x}$As /In$_{0.52}$Al$_{0.48}$As hetero-structures \cite{Das1989} where a single subband is spin-split by strong spin-orbit coupling. A spin-split Landau level gives rise to two closely spaces frequencies with similar amplitudes leading to a modulation of the dHvA amplitude given by $A \sim \textrm{cos} \pi \nu$,
where $\nu = \frac{\delta}{\hbar \omega_c}$, and $\delta$ is the energy separation between the spin-split Landau levels. Nodes in the beat pattern in dHvA will occur at half-integer values of $\nu$ ($\pm0.5, \pm 1.5,$ ....) where $A$ is zero. The total spin splitting $\delta$ can be expressed as $\delta = \delta_0 + \delta_1 \hbar \omega_c + \delta_2{(\hbar \omega_c)}^2 +$ ..., where $\delta_0$ is the zero field spitting, $\delta_1 \hbar \omega_c$ is the linear in field splitting, and $\omega_c = eB/\hbar m_c$. The higher order terms become significant at high fields. Fig. S4a show a plot of $\delta$ vs. $\hbar \omega_c$ for the fields corresponding to the nodes in dHvA oscillations in 4b (also Fig. 4a), using cyclotron mass $m_c = 0.065 m_e$ obtained from the fit of oscillations  to Lifshitz-Kosevich theory \cite{LIFSHITS1956} $\frac{\Delta \sigma_{xx}(T)}{\Delta \sigma_{xx}(0)}= \frac{\lambda(T)}{sinh \lambda(T)}$ shown in (c) using Monte Carlo technique. Here $\sigma_{xx}$ is the in-plane conductivity for magnetic field applied normal to the cleavage plane and $\lambda(T) = \frac{2 \pi k_BT}{\hbar eB}m_c$.
The extrapolation to zero field yields zero-field spin splitting $\delta_0  \approx 1.34~ \textrm{meV}$, comparable to spin splitting found in other 2DEG hetero-structures.

\section*{F. Estimating interpuddle separation}

\baselineskip24pt
Based on the material parameters in Table S2, we obtain the diffusion constant $D=v_F \ell/2 = 0.025~$m$^2$/s, and the typical interpuddle separation $a \approx140~ \textrm{nm}$. From the area $A$ of the sample $A=12.6\ \textrm{mm}^2$ we estimate the total number of puddles to be on the order $n\approx A/a^2 \approx 6\cdot10^8$.
Next we compare the absolute value of the diamagnetic response and relate it to the single puddle's response by assuming simply additive contributions of individual monodispersed puddles, $V \chi_0\approx n \chi_1$, where $V$ is the sample volume and $\chi_1$ is the average extensive susceptibility of one typical puddle. Using 
$V \approx 2.5\ \textrm{mm}^3$ we obtain $\chi_1\approx 2\cdot 10^{-22}$ at low temperatures. Relating this value to puddles' dimensions, \textit{e.g.}, radius $R$ and thickness $t$\cite{Park2013}, is complicated at high temperatures without an independent measurement of the local penetration depth, $\lambda$.  However, at low temperatures and for sufficiently large puddles we assume field exclusion which significantly simplifies the analysis, yielding $\chi_1= - 4 R^3$. Independence of this expression of the thickness $t$ is due to large demagnetization correction \cite{Hein1991}. The estimated puddle size of about $R\approx 37~ \textrm{nm}$ is comparable to the size scales of surface Dirac puddles observed by scanning tunneling microscopy \cite{Beidenkopf2011}.

\section*{G. Frequency and temperature variation of the \emph{ac} response}

\begin{figure}[b!]
\begin{center}
\includegraphics[width=16cm]{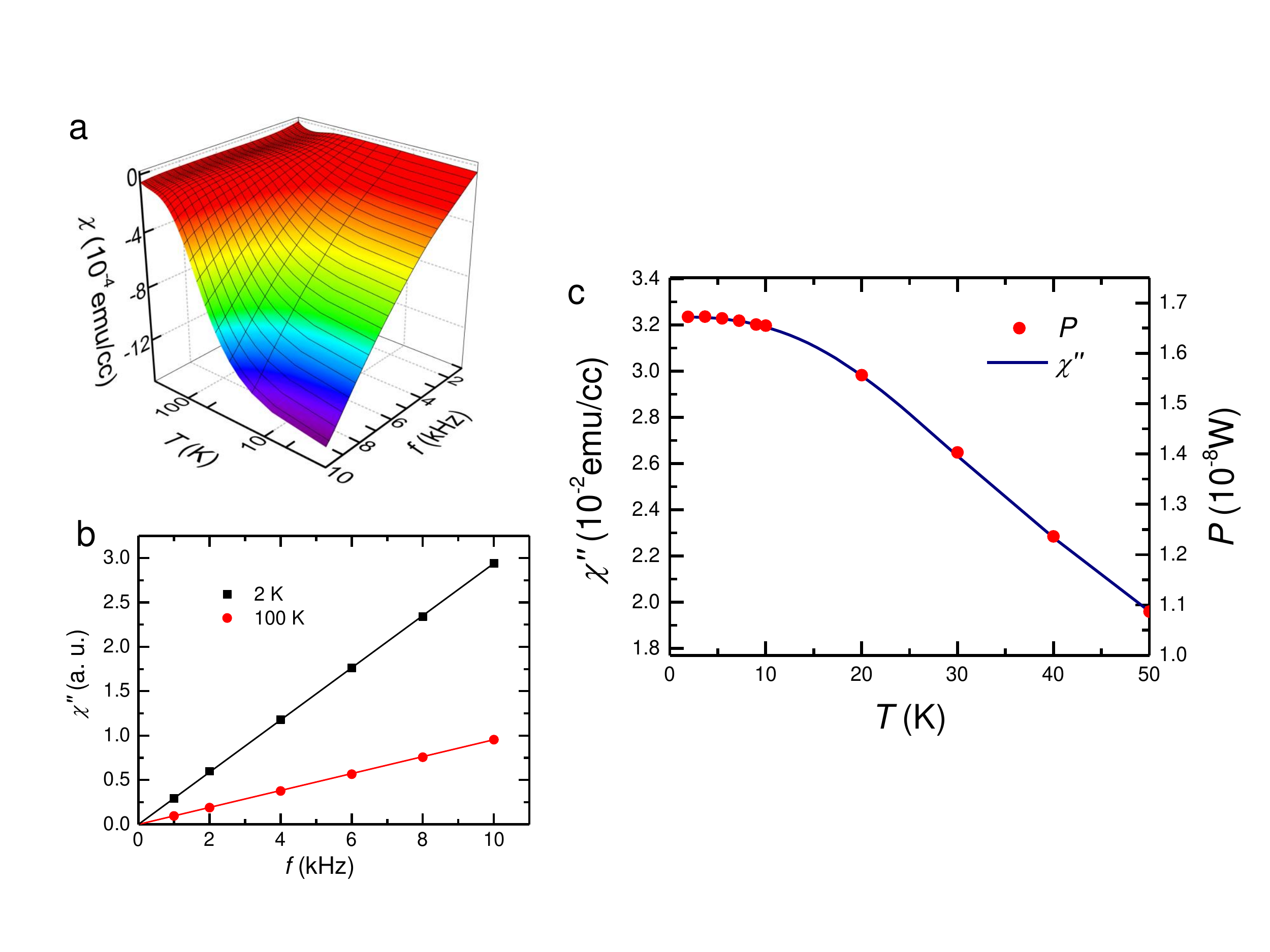}
\vspace{-10mm}
\end{center}
\caption{\label{fig:eddy}
\small (a) The in-phase component of $\chi$ is quadratic in frequency  $\chi(T) = \chi_0(T) + b(T) \omega^2$, also see Fig. S7 and main text. (b) The dissipative (out-of-phase) component of $\chi(\omega)$ is strictly frequency linear, as expected. (c) The standard eddy current mechanism is responsible for dissipation dominated by the bulk, the out-of-phase component $\chi$" is proportional to conductivity: the observed value is consistent, up to geometric factors \emph{and} closely follows the temperature dependence in-plane resistivity $\rho_{xx}$ of the \textit{bulk}, according to the standard formula for power $P = {\pi^2 (h_{ac}^2 d^2 f^2} / 2\rho_{xx} (T))$ dissipated during the \textit{ac} excitation cycles. Here  $f = \omega/2\pi $ and $d$ is the sample thickness.
 }
\end{figure}

\normalsize
\baselineskip24pt
Complex susceptibility, $\chi(\omega,T)=\chi'(\omega,T)+i \chi''(\omega,T)$, is characterized by reactive in-phase response, $\chi'$ (denoted $\chi$ in main text), but also has a dissipative out-of-phase component $\chi''$ \cite{Hein1991}.
We can readily identify the dissipative component with eddy current heating $\chi''(\omega,T)\sim \omega \sigma(T)$, where $\sigma(T)$ is the temperature dependent \emph{dc} conductivity (Fig.~\ref{fig:eddy}). Classical current fluctuations can only screen magnetic fields at high frequencies, with $\chi'(\omega\to\infty)\to -\frac{1}{4\pi}$ and $\chi'(\omega\to0)\approx -(\omega \tau)^2$, where $\tau$ is the characteristic relaxation time. For example, modeling a conductor as a simple LR-circuit we find $\tau=L/R$. Proper interpretation of finite value of $\chi'(0)<0$ requires quantum mechanics. One may, however, use purely classical phenomenology to capture diamagnetism by positing existence of macroscopic ``perfect inductor" paths, with $\tau\to\infty$. 
For superconductors we may simply think of perfect inductors as linearized Josephson elements and there exists abundant literature on phases and phase transitions of resistively shunted Josephson networks \cite{Refael2007,Giovannella1998}.
 \begin{figure}[b!]
\vspace{-57mm}
\includegraphics[width=17cm]{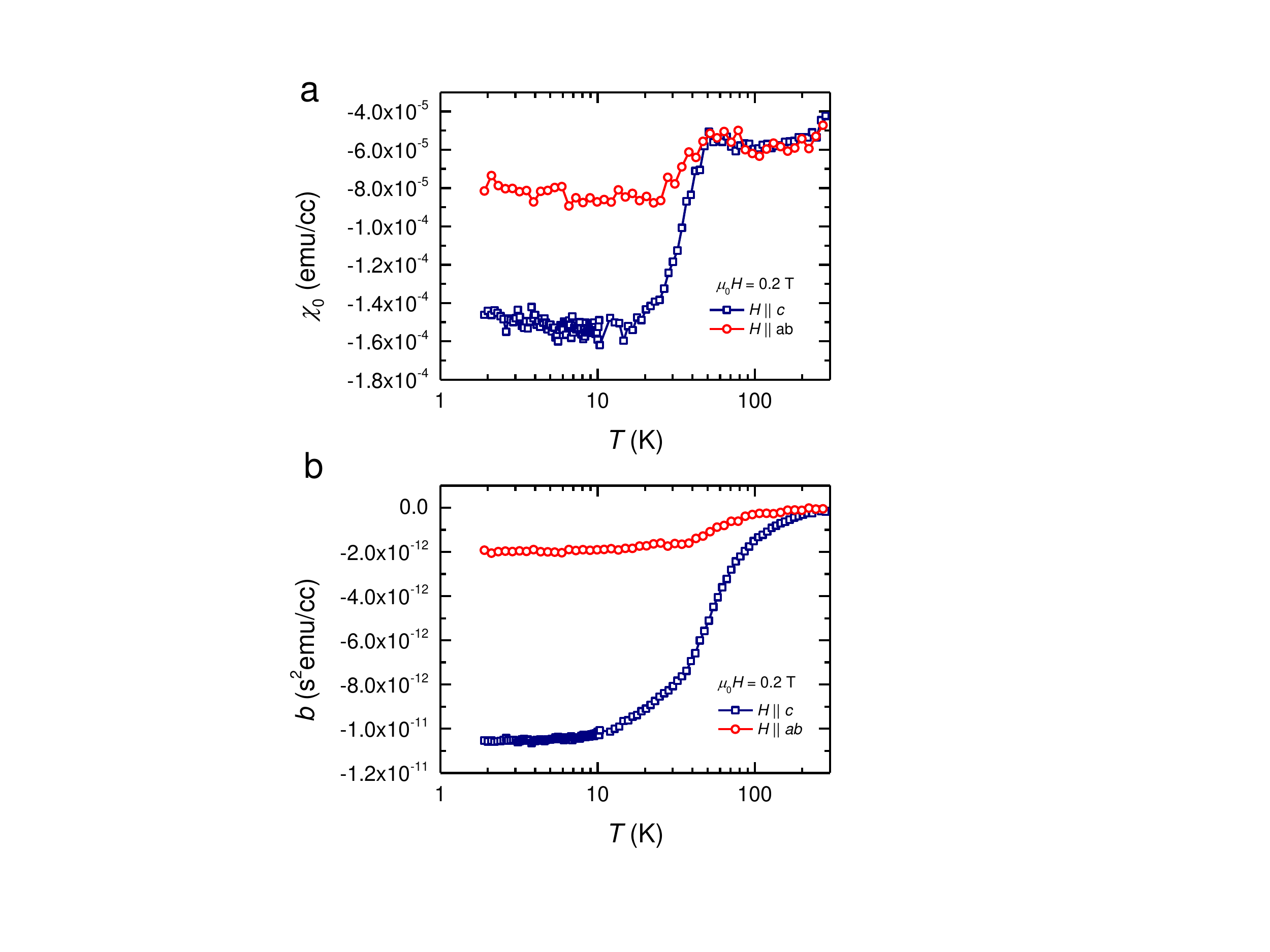}
\vspace{-15mm}
\caption{\label{fig:omega-squared}
\small (a) A sharp diamagnetic transition at $\sim 50 K$ in the zero frequency response $\chi_0(T)$ obtained from the fits to $\chi(T) = \chi_0(T) + b(T) \omega^2$ of another Sb$_2$Te$_3$ crystal.  (b) The prefactor $b(T)$ in the $\omega^2$ term is varying smoothly, dominating the total variation of $\chi$ at finite frequencies. This variation is consistent with kinetic inductance of the patchy distributed 2DEG network discussed in the main text.  }
\protect
\end{figure}

\normalsize
\baselineskip24pt
Roughly speaking, the normal phase can be thought of in a coarse grained fashion as a macroscopic $LR$ circuit, while the superconductor has $R\to0$. While this ``1-loop" phenomenology correctly captures asymptotic $\omega\to0$ limit of the two phases, it misses the low frequency correction in the superconducting phase, both $\sim\omega^2$ in $\chi'$, and, importantly $\chi''\sim\omega$, which may be thought of as the response from a finite normal fluid fraction. Various simple ``2-loop" improvements are possible to rectify this situation, \emph{e.g.,} two inductors in parallel (and only one perfect) have a simply additive response, i.e. $\chi'(\omega\to 0)=\chi_0 + b\ \omega^2 +\ldots$, with $\chi_0$ coming from perfect inductors and $b$ representative of resistively shunted elements.  We have used this phenomenology to extract $\chi_0$ in two different samples, see Fig. 3 in the main text and Fig. S7. The two measurements appear to show different amount of anisotropy for $H\parallel ab$ vs. $H\parallel c$, which we attribute to the instrumental misalignment of the latter.


\bibliography{refsMain33,booksetc}

\bibliographystyle{naturemag}




\end{document}